\documentstyle[12pt]{article}
\newcommand{\bq}{\begin{equation}}
\newcommand{\ee}{\end{equation}}

\newcommand{\fr}[2]{\frac{#1}{#2}}

\begin{document}
\pagestyle{plain}
\pagenumbering{arabic}

\begin{flushright}
{\large Preprint BUDKERINP 95-21}, March 1995
\end{flushright}

\begin{center}{\Large \bf Statistics of random quasi 1D
Hamiltonian with slowly varying parameters. Painlev\'{e}
again.}\\

\vspace{0.5cm}

{\bf  P.G.Silvestrov}\\
Budker Institute of Nuclear Physics, 630090 Novosibirsk,
Russia

\vspace{0.5cm}

\end{center}
\begin{abstract}

The statistics of random band--matrices with width and
strength of the band slowly varying along the diagonal is
considered. The Dyson equation for the averaged Green
function close to the edge of spectrum is reduced to the
Painlev\'{e} I equation. The analytical properties of the
Green function allow to fix the solution of this equation.
The former appears to be the same as that arose within the
random--matrix regularization of 2d-gravity.

\end{abstract}


{\bf 1.} The random band--matrices are now usually
considered as the theory of quasi 1D systems. For example
it is thought that band matrices may adequately depict the
properties of electrons in thin
wires \cite{Mirlin} (see also \cite{Efetov,Pichard}).

In this note we would like to consider the natural
extension of the usual band--matrix model. We will
investigate the statistical features of quasi--banded
matrices with the parameters of the band (width,
strength of interaction etc.) slowly varying along the
diagonal.

For application to condensed matter physics or energy
level statistics in complex quantum systems it may be
also useful to add some regular diagonal part to the random
Hamiltonian (see \cite{Chiricov}). Thus we deal with the
Hamiltonian
\bq\label{eq:reg}
H = H^0 +V \,\,\, , \,\,\, H^0_{ij}= h(i) \delta_{ij}
\,\,\, ,
\ee
where $h(i)$ is some smooth monotonic function and the
elements of random matrix $V_{ij}$
are Gaussian distributed.

As usual, due to the Whick theorem, the Gaussian ensemble
may be defined by the second moment:
\bq\label{eq:band}
\overline{V_{ij}V_{mn}} = \fr{F(i,j)}{b}
(\delta_{jm}\delta_{in} +
 \delta_{jn}\delta_{im}) \,\, .
\ee
Here $F(i,j)=F(j,i)$ and $b\gg 1$ is the typical width of
the band. We are interested in the quasi band case.
Therefore we suppose that $F(i,j)$ decreases rather rapidly
when $|i-j|>b$, but is a very slow smooth function of
$i+j$. In particular the sum over $j$
\bq\label{eq:W}
\fr{1}{b} \sum_j F(i,j) = W(i) \sim 1
\ee
is a very smooth function of $i$, as well as the regular
function $h(i)$,
\bq\label{eq:d}
b \fr{dW}{di} \sim b \fr{dh}{di} \sim \fr{1}{d} \ll 1 \,\,
{}.
\ee

The correlator (\ref{eq:band}) describes the Gaussian
ensemble of real symmetric matrices. One can also consider
the ensemble of general Hermitean quasi band--matrices. To
this end it is enough to withdraw the second term in
brackets in (\ref{eq:band}). All the results obtained in
this paper are relevant for both ensembles.

{\bf 2.} The "physical quantity" we would like to consider
is the Green's function $G=(E-H)^{-1}$. As follows from
(\ref{eq:band})
\bq\label{eq:G}
\overline{G_{ij}} = G(i) \delta_{ij}
\,\, .
\ee
The formal expansion in a series in $V/E$ allows to write
down the Dyson--type equation:
\begin{eqnarray}\label{eq:ser}
\overline{G_{ij}} &=& \fr{1}{E-h(i)}\left\{ \delta_{ij}+
\overline{ \sum_{n=1}^{\infty}
\left(V\fr{1}{E-H^0}\right)^n_{ij}} \right\} = \\
&=& \fr{1}{E-h(i)} \left\{ \delta_{ij} +
\sum_k
\fr{F(i,k)}{b} (
\overline{G_{kk} G_{ij}} + \overline{G_{ki} G_{kj}}
)\right\} \,\,
. \nonumber
\end{eqnarray}
Here we have performed Whick
contraction of the first
element in each $(V)^n$ with any other and then resumed
the series. In order to simplify this {\it exact} equation
it is easy to note that the second term in brackets is
small like $1/b$ and in the leading approximation over
$1/b$ one can also decouple the average of two $G$-s in the
first term. Thus
\bq\label{eq:GG}
G(i) = \fr{1}{E-h(i)}\left( 1 + \sum_k
\fr{F(i,k)}{b} G(k) G(i) \right) \,\, .
\ee
Our derivation of this formula in fact repeats the
calculation of Green function for full $N\times N$ random
matrices, which was done many years ago
\cite{Pastur,Verbashot} (see also in \cite{Brezin} the same
proof for pure band matrices).

Due to (\ref{eq:GG}) only closed chains contribute to each
$(V^n)$ in (\ref{eq:ser}). In terms of dual Feynman graphs
each $(V^n)_{ii}$ may be thought as $n$-vertex polygon,
while the averaging via Whick contractions may be
considered as a self gluing of the edges of this polygon.
Within this language in (\ref{eq:GG}) we have summed up {\it
exactly} all the planar (spherical) graphs. The nonplanar
contributions lead to corrections to $G$ of the order
of $\sim 1/b$ for real symmetric matrices and $\sim 1/b^2$
for Hermitean matrices \cite{t'Hooft}. Nevertheless the
equation (\ref{eq:GG}) is {\it exact} in $1/d$
(\ref{eq:d}).

It is easy to expand the equation (\ref{eq:ser}) in the
series in $1/d$. Up to $1/d^2$
\bq\label{eq:ser1}
G(i)= \fr{1}{E-h(i)}\left( 1 +W(i)G(i)^2 +
G(i) (AG'
+BG'') \right) \,\,  .
\ee
Here prime means the derivative with respect to $X=i/b$ and
\bq\label{eq:AB}
A(i) = \sum_k \fr{F(i,k)}{b} \left( \fr{k-i}{b} \right)
\sim \fr{1}{d} \,\,\, , \,\,\, B(i) =\fr{1}{2} \sum_k
\fr{F(i,k)}{b} \left( \fr{k-i}{b} \right)^2
\sim 1 \,\, .
\ee

In the leading approximation the solution of
(\ref{eq:ser1}) reads:
\bq\label{eq:G0}
G^0(i)= \fr{1}{2W(i)} \left( E-h(i)- \sqrt{(E-h(i))^2
-4W(i)} \right)
\,\, .
\ee
The imaginary part of $G^0$ at $E=E-i0$ reproduces the
usual in matrix models semicircle density.
In general the $1/d^2$ corrections to
(\ref{eq:G0}) may be found from (\ref{eq:ser1}) by
iteration. Due to singularity of $G^0$ this corrections,
which are proportional to $G'$ and $G''$, spreads to
infinity when $E=h(i)\pm 2\sqrt{W(i)}$. Thus the
nonperturbative
treatment of equations (\ref{eq:ser}) or (\ref{eq:ser1}) is
necessary near the edge point.

The use of expansion (\ref{eq:ser1}) assumes that
both functions $h(i)$ and $W(i)$ are equally slow
(\ref{eq:d}). The smoothness of $W$ is necessary because
we want to consider the almost banded matrices, but one can
sufficiently weaken the restriction for $h(i)$. In this
case the solution of
equation (\ref{eq:GG}) in the Breit--Wigner form is easy to
found
\begin{eqnarray}\label{Breit}
G^0(j)  = \fr{1}{E- h(j) -i\gamma(j)} \,\,\,\,\,\, &,&
\,\,\,\,\,\,
\gamma(j)= \pi F(j,j) \left( b \fr{dh}{dj} \right)^{-1}
\,\,\, , \\
\fr{1}{b} \ll \left| \fr{dh}{dj}\right| \ll \fr{1}{\sqrt{b}} \,\,\,\, &.&
\nonumber
\end{eqnarray}
Here the upper bound for $dh/dj$ ensures the smallness of
the regular energy intervals $h(j+1)-h(j)$ compared to the
hopping matrix elements $V_{ij}$ (\ref{eq:band}).

{\bf 3.} Turning back to the slow $h(i)$ case let us
introduce the new "scaling" variables
\begin{eqnarray}\label{eq:scale}
E-h(0) &=& 2 \sqrt{W(0)} + \fr{\varepsilon}{d^{4/5}} \,\, ,
\nonumber \\
X = i/b &=&  d^{1/5} x \,\,\, , \,\,\, G =
\fr{1}{\sqrt{W(0)}} + \fr{y(x)}{d^{2/5}} \,\, ,  \\
W(X) &=& W(0) + X\fr{dW}{dX} \equiv W(0) + \fr{X}{d} \,\, ,
\nonumber \\
h(X)&=& h(0) +\alpha \fr{X}{d} \,\,\, , \,\,\, \alpha \sim
1 \,\,\, . \nonumber
\end{eqnarray}
Here we have also made the explicit definition of the large
parameter $d$. Substitution of (\ref{eq:scale}) into
(\ref{eq:ser1}) leads to
\bq\label{eq:Penleve}
By'' + W^{3/2} y^2 = \varepsilon - \left(
\fr{1}{\sqrt{W}}+\alpha \right) x
\,\,\, .
\ee
Here $W=W(0)$. Due to (\ref{eq:AB}) the term $\sim AG'$ in
(\ref{eq:ser1}) leads to negligible correction of the order
of $\sim 1/d^{4/5}$. As well the higher derivatives of
$G$, which we have neglected in (\ref{eq:ser1}), contribute
like some powers of $1/d$.

The equation (\ref{eq:Penleve}) is the famous Painlev\'{e}
I equation. Parameters $B, W$ and $\alpha$ which still
alive are of the order of $1$ and may be dropped out by
trivial rescaling.
Being a part of Green the function
$y=y_{\varepsilon}(x)$,
have to reproduce its analytical features. Thus the only
singularity of $y_{\varepsilon}(x)$ is the cut along
the axis $Im\varepsilon=0$. We have defined $t$
(\ref{eq:scale}) as a real variable. On the other hand it
is natural to consider $y_{\varepsilon}(x)$ as an
analytical function of one complex variable $x'=x -
\fr{\sqrt{W}}{1+\alpha \sqrt{W}}\varepsilon$. Now the
function
$y(x')$ should
not have
singularities in the whole upper half plane.

It is amazing that our model is not the first model of
random matrices whose scaling limit is described by the
Painlev\'{e} equation. A few years ago the random matrix
models helped to solve exactly the $2d$ quantum gravity
\cite{BKGMDS}. It was shown that the second derivative of
partition function of $2d$ gravity with respect to the
cosmological constant is the solution of Painlev\'{e}
equation. The unique solution of this equation, which is
realized in Euclidean $2d$--gravity was found in
\cite{DmeIts}. Each solution of eq. (\ref{eq:Penleve}) has
an infinite set of second order poles on the complex plane
$y\sim (x-x_i)^{-2}$ and no other singularities. As it was
pointed out by F.David \cite{DmeIts} (and as was known for
mathematicians for many years) only one solution may have no
poles in the whole half plane. This is the so called
"triply truncated solution", which may have only a finite
number of poles in the sector $\fr{-2\pi}{5} < Arg(x) <
\fr{6\pi}{5}$ (and still has an infinite set of poles
within the rest $\fr{2\pi}{5}$). Just this solution
is realized in $2d$--gravity if the matrix model is
regularized via analytical continuation. Moreover only this
unique solution can reproduce the analytical features of the
Green function in our model (\ref{eq:band}).

The most informative quantity, which may be found from the
averaged Green function is the single particle density
\bq\label{eq:rho}
\rho (i,E)= \fr{1}{\pi} ImG(i,E-i0) \sim Im \,
y(x-\fr{\sqrt{W}}{2} \varepsilon) \,\, .
\ee
It is easily seen that for real $\varepsilon$ all real
solutions of (\ref{eq:Penleve}) are unstable (have the
poles at real $x$ axis). The complex solutions, in
accordance with (\ref{eq:G0}), have all the same asymptotics
$Im\, y(x\rightarrow \pm \infty )= -
\sqrt{\varepsilon-(\fr{1}{\sqrt{W}}+\alpha x)}$.
Nevertheless the
nonperturbative imaginary part of $y$ at negative $x$
allows to distinguish the triply truncated solution among
the others \cite{DmeIts,David2}
\begin{eqnarray}\label{Imy}
Im\, y(x\rightarrow -\infty ) &=&
\fr{\sqrt{3\sqrt{2}}}{4\sqrt{\pi}}
\fr{B^{1/4}(1+\alpha\sqrt{W})^{3/8}}{W^{9/8}}
\left(\fr{\varepsilon\sqrt{W}}{1+\alpha\sqrt{W}} -
x\right)^{-1/8}
\nonumber \\
& & \exp
\left\{ -\fr{4\sqrt{2}}{5\sqrt{B}} W^{1/4}
(1+\alpha\sqrt{W})^{1/4} \left(
\fr{\sqrt{W}}{2} \varepsilon -x\right)^{5/4} \right\} \,\,
{}.
\end{eqnarray}

Thus we see that close to the edge the single particle
density behaves like
\bq\label{eq:rhod}
\rho (i,E) = \fr{1}{d^{2/5}} f\left( d^{4/5}
 (E-E_0) \right) \,\,\, ,
 \,\,\,
 \ee
 where $f$ is a smooth imaginary part of the triply
truncated solution. This result accounts exactly for the
series of most singular corrections over $1/d^n$
(\ref{eq:d},\ref{eq:scale}).

Still we have not considered the corrections of the order
of $1/b$. These corrections accounts for
the finite width of the band (\ref{eq:band}) and also can
smooth out the
singularity of the zero order result (\ref{eq:G0}). The
accurate description of $1/b$ effects is beyond of
the main subject of this paper. Here we give only the
result and postpone discussion for a separate publication
\cite{me}. For the pure band--matrix (i.e. if
$F(i,j)\equiv F(|i-j|) $ and $h(i)$ is negligible small)
the density of states close to
end--point reads
\bq\label{eq:rhob}
\rho (E) =\fr{1}{b^{2/5}} \phi (b^{4/5} (E-E_0))
\ee
with some smooth $\phi$. So we can conclude, that the
scaling (\ref{eq:scale},\ref{eq:Penleve}) should not be
disturbed by the $1/b$ corrections until $d \ll b$.

{\bf 4.} As we have seen (\ref{eq:scale},\ref{eq:Penleve})
the variation of band width $W(i)$ and shifting of the
center of zone $h(i)$ play the same role at the edge.
Therefore in the following two sections we suppose that
$h=0$.

Up to now we have supposed (\ref{eq:scale}) that
the edge of spectrum $E_e =h(i)+2\sqrt{W(i)}$ is a smooth
monotonic function of $i$.
On the other hand it seems to be interesting to investigate
also the statistical properties of the model
(\ref{eq:band}) near the bulge of $E_e$. The set of scaling
variables suitable for this
case reads
\begin{eqnarray}\label{eq:scale1}
W(X) &=& W(0) - \fr{X^2}{2d^2} \,\,\, , \,\,\, X = i/b =
\fr{B^{1/3}}{W^{5/6}} d^{1/3} x\,\,\, , \\
E &=& 2 \sqrt{W(0)} + \fr{B^{2/3}}{W^{5/6}}
\fr{\varepsilon}{d^{4/3}} \,\,\, , \,\,\, G =
\fr{1}{\sqrt{W(0)}} + \fr{B^{1/3}}{W^{7/6}}
\fr{y(x)}{d^{2/3}} \,\, .
\nonumber
\end{eqnarray}
Here again $W=W(0)$ and $B$ are of the order of $1$. Also
here we give a new definition of large $d$. Simply
substituting (\ref{eq:scale1}) into the same equation
(\ref{eq:ser1}) one gets
\bq\label{eq:Penleve1}
y'' +y^2 =\varepsilon +x^2
\,\,\, .
\ee
Like in monotonic case (\ref{eq:Penleve}) the first
derivative $y'$ and all higher derivatives are
suppressed by some powers of $1/d$. The asymptotic solution
of this equation reads
\bq\label{eq:asym}
y = -\sqrt{\varepsilon +x^2} + \dots
\ee
in accordance with (\ref{eq:G0}). For large negative
$\varepsilon$ in vicinity of $\varepsilon +x^2=0$ the
equation (\ref{eq:Penleve1}) reduces to the Painlev\'{e} I
and the singularity of (\ref{eq:asym}) is smoothed out via
its triply truncated solution. Thus the only new problem
will be to find $y$ at $\varepsilon \sim 1$.

Generally speaking the equation (\ref{eq:Penleve1}) is not
of the Painlev\'{e} type. Its solutions also has poles on
the complex plane, but each pole turns out to be also the
edge of logarithmic cut. We are interested in the set of
real $x$ solutions $y_{\varepsilon}(x)$ analytical in the
whole lower half plane of complex parameter $\varepsilon$.
Let us write down explicitly the real and imaginary parts
of (\ref{eq:Penleve1})
\begin{eqnarray}\label{eq:system}
\varepsilon \rightarrow \varepsilon -i\lambda , y=u+iv
\,\,\, , \nonumber \\
\left\{ \begin{array}{ll}
         u'' + u^2 -v^2 = \varepsilon + x^2 \\
         v'' + 2uv = - \lambda   \,\,\,\,\,\,\, .
         \end{array}
         \right.
\end{eqnarray}
For real energy ($\lambda =0$) the latter equation may be
thought as the Schr\"{o}dinger equation with $v$ being the
wave function, $(-u)$ -- the potential and eigenvalue equal
to zero.
Moreover it is immediately seen from (\ref{eq:asym}) that
this Schr\"{o}dinger equation could not have any
localized solution at sufficiently large positive
$\varepsilon$. Thus the imaginary part of Green
function (or the single particle density) vanishes in our
model at large $\varepsilon$.

Our scaling equation (\ref{eq:Penleve1}) accounts only for
the most singular corrections in each order over $1/d$.
Nevertheless, one can see, that for the {\it exact}
equation (\ref{eq:ser}) the imaginary part of $G(i)$ also
vanishes at sufficiently large $E$. In fact at large $E$
the imaginary part at is small and thus it
should satisfy some linear homogeneous matrix equation. But
in general (except for the degenerate case) the linear
equation has no nontrivial solutions.

The exact solution of equation (\ref{eq:Penleve1})
(or(\ref{eq:system})) may be found only numerically.
Nevertheless one can further clarify, how the imaginary
part appears for real $\varepsilon$. Suppose that
$y=u_0(x)$ is the real solution of (\ref{eq:Penleve1}) with
smallest real $\varepsilon_0$. Then the
corresponding Schr\"{o}dinger equation at $u=u_0$ also
should have the localized zero mode $\psi_0$~:
\begin{eqnarray}\label{eq:zero}
\left\{ \begin{array}{ll}
         u''_0 + u_0^2 = \varepsilon_0 + x^2 \\
         \left(-\fr{d^2}{dx^2} - 2u_0 \right) \psi_0 = 0
\,\,\,\,\,\,\, .
         \end{array}
         \right.
\end{eqnarray}
Now it is rather simple exercise to find the solution in
vicinity of $\varepsilon_0$:
\bq\label{eq:sqrt}
\varepsilon = \varepsilon_0 +\Delta \,\,\, , \,\,\,
y= u_0 - \sqrt{ \Delta \fr{\int \psi_0 dx}{\int \psi_0^3
dx}} \, \psi_0 + O(\Delta) \,\,\, .
\ee
So we see that the square root singularity of the
zero order result (\ref{eq:G0}) which seemed to be
smoothed out by the solution of differential equation
(\ref{eq:Penleve1}) still alive at the very edge of
the spectrum. This result may have even experimentally
measurable consequences. One can introduce the density of
eigenstates (see (\ref{eq:rho}))
\bq\label{eq:rhoE}
\rho(E)= \sum_i \rho(i,E) \,\,\, .
\ee
This quantity is well defined because we are working at the
point of maximum thickness of the wire (band). At large
negative $\varepsilon$ the asymptotics of $\rho(E)$ may be
found from (\ref{eq:asym})
\bq\label{eq:rhoas}
-\varepsilon \gg 1 \,\,\, , \,\,\, \rho(\varepsilon) \sim
(\varepsilon_0 - \varepsilon)^{3/2} \,\,\, .
\ee
On the other hand close to the edge
\bq\label{eq:rhoedg}
0 < \varepsilon_0 - \varepsilon \ll 1 \,\,\, , \,\,\,
\rho(\varepsilon) \sim \sqrt{\varepsilon_0 - \varepsilon}
\,\,\, .
\ee
Thus the "smooth" exact result appears to be even more
singular than the zero order approximation
(\ref{eq:rhoas}).

Of course the vanishing of $Im\,y$ at $\varepsilon >
\varepsilon_0$ is the artifact of our approximation. Again
in this section we have neglected all the $1/b$
corrections. Comparing (\ref{eq:scale1}) and
(\ref{eq:rhob}) one can find that our results are valid if
$d \ll b^{3/5}$.

{\bf 5.} Up to now we have considered the averaging of only
the single Green function. The much more puzzling
quantities are the correlation functions of two Green
functions. In particular the correlators are sensitive to
the effect of localization. For usual band--matrices the
localization length turns out to be of the order of $\sim
b^2$ \cite{Mirlin} and can not be found without
nonperturbative treatment of $\sim 1/b$ effects.
Nevertheless near the thickening of the wire
(\ref{eq:scale1}) some sort of compulsory localization
takes place and the eigen-functions are localized due to
the "geometry" of band (wire). In this section we would
like to find the averaged value of the squared absolute
value of $G_{ij}$ at the thickening point.  In fact what we
can easily calculate is the so called "smoothed" correlator
\cite{Brezin}, valid only for sufficiently large $ImE$ (at
least as compared to the interval between the individual
levels).

Like in (\ref{eq:ser}) one can expand correlator in the
series
\bq\label{eq:serGG}
\overline{ |G_{ij}|^2 } = \fr{1}{|E|^2} \sum_{n,m}
\overline{ \left( \fr{V}{E} \right)^n_{ij}
\left( \fr{V}{E^*} \right)^m_{ji} } \,\,\,\,\, .
\ee
We are mostly interested in the case of small imaginary
part of $E$. The crucial observation is that in the planar
limit all the Whick contractions between $(V)^n$ and
$(V)^m$ may be done explicitly, while the sum over
$(V/E)^s$ which are contracted within one of the
multipliers in (\ref{eq:serGG}) may be resumed back to
$G(i)$ or $G(i)^*$ (\ref{eq:G}):
\bq\label{eq:sumGG}
\overline{ |G_{ij}|^2 } = \sum_{n=0} \sum_{i_1,\dots
,i_n}
|G|^2_i \fr{F(i,i_1)}{b} |G|^2_{i_1} \dots \fr{F(i_n,j)}{b}
|G|^2_j \equiv \sum_{n=0} \Gamma_n (i,j) \,\,\,\, ,
\ee
where $|G|^2_i \equiv |G(i)|^2$. Like the equation
(\ref{eq:GG}) this formula accounts {\it exactly} for all
$\sim 1/d$ corrections. All the $\sim 1/b$ corrections to
$|G_{ij}|^2$ have been neglected in (\ref{eq:sumGG}).
Unfortunately the price for this approximation is rather
high.  The $\sim 1/b$ corrections will blow up the
correlation function at very small $ImE$, thus preventing
the observation of the correlations in neighbor levels.

The recursion formula
for $\Gamma$-s (compare with(\ref{eq:GG},\ref{eq:ser1})) reads
\begin{eqnarray}\label{eq:rec}
\Gamma_{n+1}(i,j) &=& \sum_k |G|^2_i \fr{F(i,k)}{b}
\Gamma_{n}(k,j) = \\
&=& W|G|^2 \Gamma_{n} + A|G|^2 \Gamma_{n}'
+B|G|^2 \Gamma_{n}'' + \dots \,\,\,\, , \nonumber
\end{eqnarray}
where $A,B,|G|^2$ are taken at the point $i$ and the prime
means the derivative with respect to $X=i/b$. Now if one
uses
the scaling variables (\ref{eq:scale1}) and also introduces
the additional variable (time) instead of $n$ the
differential equation for $\Gamma (t,x,x_j)$ is easy to
write down
\bq\label{eq:diff}
t=\fr{n}{d^{2/3}} \,\,\, , \,\,\, \dot{\Gamma} = \Gamma'' + 2
Re\, y \, \Gamma \,\,\, .
\ee
Here and below we choose $W(0)=B=1$ for simplicity. It is
seen immediately from (\ref{eq:system}) that $\Gamma$
spreads to infinity if $Im \, \varepsilon \rightarrow 0$
due to zero mode of the Schr\"{o}dinger equation in the
r.h.s. of (\ref{eq:diff}).

For small $\lambda = - Im\, \varepsilon$ it is convenient
to consider the linearized version of the system
(\ref{eq:system})
\begin{eqnarray}\label{eq:systeml}
u= u_0 + f \,\, , \,\, v= v_0 + g \,\, , \,\,\, f,g \sim
\lambda
\,\,\, , \nonumber \\
\left\{ \begin{array}{ll}
         f'' + 2u_0 f = 2v_0 g \\
         g'' + 2u_0 g = - \lambda -2v_0 f  \,\,\,\,\,\,\, .
         \end{array}
         \right.
\end{eqnarray}
Here $u_0,v_0$ are the solutions of (\ref{eq:system}) for
$\lambda = 0$ and hence $v_0$ is also the zero mode of
operator $\fr{d^2}{dx^2} + 2u_0$. Thus one can write down
the consistency conditions for two equations
(\ref{eq:systeml}):
\bq\label{eq:cons}
2\int v_0^2 g = 0 \,\,\, , \,\,\, \lambda \int v_0 = -2 \int
v_0^2 f \,\,\, .
\ee
The solution of (\ref{eq:diff}) may be found as a sum over
eigenmodes of the operator:
\begin{eqnarray}\label{eq:operator}
-\fr{d^2}{dx^2} - 2u &=& -\fr{d^2}{dx^2} - 2u_0 -2f \,\,\, ,
\nonumber \\
\Gamma(t,x,x_j) &=& \fr{1}{b d^{1/3}} \sum |n\rangle \langle
n| \exp(-\epsilon_n t) \,\,\, ,
\end{eqnarray}
where $\epsilon_n$ and $|n\rangle$ are the eigenvalues and
normalized eigenmodes and the overall factor in $\Gamma$
was found from the initial condition (see
(\ref{eq:sumGG})). For
small $\lambda$ and large $t$ only the almost zero mode
($n=0$) survives in (\ref{eq:operator}), which is:
\bq\label{eq:zeromode}
|0\rangle = \fr{v_0(x)}{\sqrt{\int v_0^2}} \,\,\, ,
\epsilon_0 = -2\fr{\int v_0^2 f}{\int v_0^2} = \lambda
\fr{\int v_0}{\int v_0^2} \,\,\, .
\ee
Finally the "smoothed" \cite{Brezin} correlator reads
\bq\label{eq:result}
\overline{|G_{ij}|^2} = \fr{1}{bd\,Im\,E} \,\,\,
\fr{v_0(x_i) v_0(x_j)}{\int v_0 dx} \,\,\, , \,\,\,
\sum_j \overline{|G_{ij}|^2} = \fr{\rho(i,E)}{\pi \, Im\,
E}
 \,\,\, .
\ee
Here we use that $ \lambda = d^{4/3} Im\,E$
(\ref{eq:scale1}). We have also shown the sum rule, which
this
correlator evidently satisfy. Function $v_0(x)$ may be
found numerically as a solution of the simple system of
differential equations (\ref{eq:system}).

{\bf Ascnowledgements.} Author is thankful to
B.~V.~Chirikov, D.~V.~Savin, \\ V.~V.~Sokolov and
O.~P.~Sushkov for useful discussions.

\end{document}